\documentclass[aps, pre, twocolumn,showkeys,showpacs,amsmath,amssymb,superscriptaddress]{revtex4-1}
\usepackage{graphicx}
\usepackage{dcolumn}
\usepackage{bm}
\usepackage{epsfig}
\usepackage{subfigure}
\usepackage{color}
\usepackage{epstopdf}

\begin{document}

\title{Lattice Boltzmann Model for Numerical Relativity}
%\shorttitle{Title} %Insert here a short version of the title if it exceeds 70 characters

\author{E. Ilseven} \email{eilseven@student.ethz.ch} \affiliation{ ETH
  Z\"urich, Computational Physics for Engineering Materials, Institute
  for Building Materials, Wolfgang-Pauli-Strasse 27, HIT, CH-8093 Z\"urich
  (Switzerland)}

\author{M. Mendoza} \email{mmendoza@ethz.ch} \affiliation{  ETH
  Z\"urich, Computational Physics for Engineering Materials, Institute
  for Building Materials, Wolfgang-Pauli-Strasse 27, HIT, CH-8093 Z\"urich
  (Switzerland)}

%\pacs{47.11.-j}{Computational methods in fluid dynamics}
%\pacs{47.75.+f}{Relativistic fluid dynamics}
%\pacs{47.20.-k}{Flow instabilities}

\begin{abstract}
In the Bona-Masso formulation, Einstein equations are written as a set of flux conservative first order hyperbolic equations that resemble fluid dynamics equations. Based on this formulation, we construct a lattice Boltzmann model for Numerical Relativity. Our model is validated with well-established tests, showing good agreement with analytical solutions. Furthermore, we show that by increasing the relaxation time, we gain stability at the cost of losing accuracy, and by decreasing the lattice spacings while keeping a constant numerical diffusivity, the accuracy and stability of our simulations improves. Finally, in order to show the potential of our approach a linear scaling law for parallelisation with respect to number of CPU cores is demonstrated. Our model represents the first step in using lattice kinetic theory to solve gravitational problems.
\end{abstract}
\date{\today}
%\pacs{47.11.-j, 47.75.+f, 47.20.-k}
%\pacs{47.11.-j, 02.40.-k, 95.30.Sf}

\maketitle
\section{Introduction}
\label{Introduction}
%Computational fluid dynamics is a well-studied branch of physics, which has wide applications from engineering problems to astrophysics. There are several numerical methods to solve the fluid dynamics equations, which are some of the many conservation equations in physics. Another branch of physics is\\\linebreak 

General Relativity was introduced by Albert Einstein in 1915, and became the first geometric theory of gravitation, including 10 non-linear second order differential equations. Due to its complexity, few analytical solutions have been found till today, and therefore, numerical methods have played an important role in the last decades. Solving the Einstein equations numerically is often called Numerical Relativity.

Many different systems such as gauge waves, gravitational waves and Schwarzschild black hole systems are solved using Numerical Relativity \cite{pretorius2005evolution,centrella2010black,gomez1998stable}. With the 2005 breakthroughs, the binary Schwarzschild black holes and merger simulations last long enough to extract gravitational waves and study their forms \cite{campanelli2006accurate, baker2006gravitational}. Most popular methods for solving the equations are based on finite difference schemes combined with mathematical or numerical correction techniques \cite{anninos1995three,bona1997first}. While there is a big effort in more complicated phenomena, there is also much work done in solving numerical problems encountered during the numerical simulations. Numerical Relativity has mainly two problems besides stability and accuracy: insufficient memory, and complicated geometries that lead to emerging singularities \cite{anninos1995three}. 
%As there are 12 equations to solve simultaneously, the available memory allows computations only in one or two dimensions. The memory of a normal computer is insufficient in three dimensions and errors due to boundary conditions affect the system. 
Such problems, especially in highly curved spacetimes, are tackled by mathematical corrections, such as conformal and isometric mappings \cite{anninos1995three}, or by numerical methods, such as exponentially growing lattice spacings or adaptive mesh refinements \cite{brugmann1996adaptive}. The complicated geometries and singularities also lead to highly curved spacetimes and some simulations are expected to crash at a certain time \cite{brugmann1996adaptive}. 

There are several mathematical formulations of Einstein equations \cite{centrella2010black}. In particular, the $3+1$ decomposition is the most convenient mathematical model for numerical calculations and it sets the basis for different formalisms. Arnowit-Deser-Misner (ADM) and BSSNOK (Baumgarte, Shapiro, Shibata, Nakamura, Oohara, Kojima) formalisms have proven to be the most stable ones at the present \cite{campanelli2006accurate, baker2006gravitational,alcubierre2008introduction}. In this work, we use the Bona-Masso formulation of Einstein equations, since it consists in first order hyperbolic conservation equations for the geometric variables \cite{bona1997first,bona1995new}, resembling fluid dynamics equations, and therefore, making suitable the use of fluid dynamics solvers.

In this paper, we propose for the first time a lattice Boltzmann model to solve Einstein equations. Usually, the lattice Boltzmann method allows us to solve the Navier-Stokes equations (or any conservation law) to an accuracy that depends on the Knudsen number (roughly defined as the ratio of lattice spacing to system size) \cite{wolf2000lattice}. This method has been successfully used to study many physical systems using a fraction of computational time in comparison with other numerical methods \cite{mendoza2010three,succi1993lattice,mendoza2010fast,mendoza2014kinetic,mendoza2013flow}. Here, we extend the wide applicability of the lattice Boltzmann method to gravitation. Furthermore, we investigate the performance of the model with the following well-established tests reported in Ref. \cite{alcubierre2004towards}. \textit{Expansion of a flat universe:} We simulate an expanding flat universe and compare the results with the expected analytical solution. This example uses an ideal fluid energy-momentum tensor and shows that the energy-momentum tensor is coupled correctly to the rest of the model variables. \textit{Gauge wave:} We test the ability of the scheme to propagate a gauge wave. The gauge wave is achieved through a coordinate transformation, which does not change the physics at hand. \textit{Linear wave:} We validate the model propagating the amplitude and phase of a gravitational wave, i.e. a spatial transverse wave propagating in $x$-direction and time. \textit{Numerical diffusivity:} We investigate the role of numerical diffusivity in our simulation errors. While it is normally added artificially in other methods through an elliptic equation \cite{anninos1995three}, it is present in the lattice Boltzmann model naturally. \textit{Parallelization:} We show that our lattice Boltzmann model is well suited for parallel computing. The improvement in the computation duration for a gauge wave simulation is also shown.
%\item \textit{Stability tests:} While testing the code for the above mentioned simulation, we are going to also check the stability of our code according to different relaxation times, which is a characteristic parameter in Lattice Boltzmann.
%\\\linebreak 
% In order to find the right application of boundaries in LBM, one has to understand the boundaries in NR very well which is beyond the scope of this research.

Compared to common methods in Numerical Relativity, such as finite differences, lattice Boltzmann methods have extra properties that may lead to improvements or short comings, which still need to be tested. Some of them can be listed as follow: The numerical viscosity in Numerical Relativity is an extra term introduced in finite difference methods (see Ref. \cite{anninos1995three}) to add numerical stability, while it is naturally present in the lattice Boltzmann method; our model uses a lattice that offers more isotropy and accurate results when it comes to solving hyperbolic equations; and finally, for simple cases, the lattice Boltzmann method sets time step equal to the lattice spacing, i.e. space and time scale linearly. Therefore, the present work is just the beginning of a new way of solving Einstein equation and opens up the door for further research in Numerical Relativity.

This paper first gives an introduction to Numerical Relativity and the theory of the specific formulation of the Einstein equations that we have chosen. The lattice Boltzmann method is introduced in the next section. Afterwards, the validation tests are described and their results using our numerical method are analysed. In the last section, we summarise our work and discuss possible outlooks.

\section{Theory}
\label{MNM}

Einstein equations are $10$ coupled second order nonlinear differential equations which, in their most compact form, are given by
\begin{equation}
R_{\alpha \beta} - \frac{1}{2}g_{\alpha \beta}R = T_{\alpha \beta},
\label{einstein}
\end{equation}
where $T_{\alpha \beta}$ is the energy-momentum tensor, $g_{\alpha \beta}$ the four metric and $R_{\alpha \beta}$ the Ricci tensor defined as
\begin{equation}
R_{\alpha \beta} = \partial_{\rho}\Gamma^\rho_{\alpha \beta} - \partial_{\beta}\Gamma^\rho_{\alpha \rho} + \Gamma^\rho_{\lambda \rho}\Gamma^\lambda_{\alpha \beta} - \Gamma^\rho_{\lambda \beta}\Gamma^\lambda_{\alpha \rho},
\end{equation}
with $\Gamma^\rho_{\alpha \beta}$ the Christoffel symbols,
\begin{equation}
\Gamma^\rho_{\alpha \beta} = \frac{1}{2}g^{\rho \lambda}(\partial_\alpha g_{\beta \lambda} + \partial_\beta g_{\alpha \lambda} - \partial_\lambda g_{\alpha \beta}).
\end{equation}
We use Einstein summation convention throughout the paper and the greek letters indicate summation over four dimensions, while the latin letters indicate summation over the three spatial components. Einstein equations, Eq.~\eqref{einstein}, in their expanded forms possess very complex shapes and several formalisms are used to ease their solutions. 

\subsection{3+1 Arnowit Deser Misner formalism}

The ADM formalism of Einstein equations was first introduced through Hamiltonian framework of General Relativity \cite{alcubierre2008introduction,arnowitt1962dynamics}. Here, we obtain the ADM formalism with a slightly different construction, as described below. The $3+1$ decomposition of Einstein equations lets us find a solution for the four-metric in different hyper-spaces, and is given as follows:
\begin{equation}
ds^2 = \alpha dt^2 - \gamma_{ij}(dx^i + \beta^idt)(dx^j + \beta^jdt) \quad ,
\end{equation}
where $\alpha$ and $\beta_i$ are describing the evolution of hyper-spaces in the four dimensional space, and the hyper-spaces (also called slices) are associated with a certain three-metric $\gamma_{ij}$. The construction of these geometric objects is given by the following scheme: we construct three dimensional spaces with the three-metric $\gamma_{ij}$ and find the normal vector to these spaces with components $\alpha$ and $\beta_i$. At every time step we need a variable that describes the evolution of $\gamma_{ij}$, which is given by the extrinsic curvature $K_{ij}$, which is the projection of the curvature to the normal vector onto the three spaces. Due to the symmetry of $\gamma_{ij}$ (thus $K_{ij}$) we arrive to $6$ differential equations describing the evolution of the system. Note that other $4$ degrees of freedom are still left, as we are dealing with $10$ differential equations. These degrees of freedom are covered by the constraint equations, which are energy and momentum conservation. Finally the ADM equations and constraints are given by the following equations:
\begin{equation}
(\partial_t - L_{\beta}) \gamma_{ij} = -2 \alpha K_{ij},
\end{equation}
\begin{multline}
(\partial_t - L_{\beta}) K_{ij} = -\alpha_{i;j} + \\ \alpha(R^{(3)}_{ij}-2K_{im}K^m_j + \text{tr}(K)K_{ij} - R^{(4)}_{ij}),
\label{secfev}
\end{multline}
\begin{equation}
R^{(3)} - \text{tr}(K^2) + \text{tr}(K)^2 - 2\alpha^2 G^{00} = 0,
\end{equation}
\begin{equation}
K^j_{i;j} - \partial_i(\text{tr}(K)) - \alpha G^0_i = 0,
\end{equation}
where $L_{\beta}$ corresponds to the Lie derivative along the vector $\beta$ and $G_{\alpha \beta}$ to the energy momentum tensor, and $R^{(4)}_{ij}$ is the projection of the four Ricci tensor on the three dimensional spaces. These equations correspond to one representation of the Einstein equations. As we will see later, there can be other formulations. Indeed, all \textit{physical} solutions of the equations match, even if the formulations may not have the same \textit{mathematical} structure. The physical solutions are known to be the solutions belonging to the ``constrained space". However, the mathematical differences may increase or decrease the stability of numerical schemes. These two equations can be evolved in two different ways. The first one is letting the system evolve freely and monitor the constraints (``free evolution"), and the second one, is solving the constraint equations for each time step for some of the variables and make sure that they are fulfilled (``constrained evolution"). It must be pointed out that the $3+1$ decomposition does not lead to an evolution equation for the slicing $\alpha$ and the shift $\beta_i$. So we have four additional degrees of freedom, which we can exploit to improve stability in different systems. Thus, we can formulate
\begin{equation}
(\partial_t - L_{\beta}) \alpha = -\alpha^2 Q,
\end{equation}
\begin{equation}
(\partial_t - L_{\beta}) \beta_i = -\alpha^2 Q_i.
\end{equation}
where $Q$ and $Q_i$ are gauge functions.

\subsubsection{Bona-Masso formulation}

While there are many ways to solve Einstein equations numerically, the following two methods are the most popular ones: ADM and BSSNOK (Baumgarte, Shapiro, Shibata, Nakamura, Oohara, Kojima) formalisms \cite{alcubierre2008introduction}. While ADM tries to solve the equations directly, BSSNOK makes use of conformal mappings and increases the stability of the evolution as conformal variables are evolved. Both formalisms are solved mainly by finite difference methods. In this work, we will use a third method known as Bona-Masso formalism \cite{bona1997first,bona1995new,bona2005geometrically}. It consists in a set of first order flux conservative hyperbolic equations equivalent to Einstein equations. We have seen that the Einstein equations, within the ADM formalism, are first order in time derivatives but second order in space derivatives. Thus, the Bona-Masso formalism introduces the following space derivatives to obtain first order differential equations:
\begin{equation}
A_k = \partial_k \ln(\alpha), \;\;\;\; B_k^i = \partial_k \beta^i/2, \;\;\;\; D_{kij} = \partial_k \gamma_{ij}/2.
\end{equation}

There are two evolution systems in this formalism \cite{bona1997first}. The first one is the ``Ricci evolution system", which consists in the above formulated ADM equations, and the second one is the ``Einstein evolution system". In the latter one, the four Ricci tensor is replaced by the energy-momentum tensor (using Eq.~\eqref{einstein}). Thus in the Einstein system the energy constraint is already solved by the evolution equations. The Bona-Masso formulation introduces the following variable similar to the energy constraint:
\begin{equation}
V_{k} = D_{kr}^{\;\;r} - D^r_{rk}.
\end{equation}
The quantity $V_i$ allows us to implement the momentum conservation in its evolution equations such that we do not need to solve the momentum constraint separately, i.e. the momentum conservation is solved by the equations for $V_i$ during the evolution. Now we can take a look at the Einstein equations through the Bona-Masso formalism. As mentioned in the introduction, we are considering zero shift cases, which means $\beta_i = 0, Q_i = 0 \implies \beta_i(t) = 0$. The expressions include the shift vector only for general purposes.

\subsection{Bona-Masso evolution equations}

We are considering the Einstein evolution system. The Bona-Masso equations are given by the following equations:
\\\linebreak\textit{\underline{Slicing:}}
\begin{equation}
\partial_t \alpha =  \alpha (\beta^r A_r - \alpha Q),
\label{alfan}
\end{equation}
\\\linebreak\textit{\underline{Three metric:}}
\begin{equation}
\partial_t \gamma_{ij} = 2\beta^r D_{kij} - 2\alpha (K_{ij}-s_{ij}),
\end{equation}
\begin{equation}
s_{ij}  =Ê\frac{B_{ij}+B_{ji}}{\alpha},
\end{equation}
\\\linebreak\textit{\underline{Extrinsic Curvature:}}
\begin{equation}
\partial_t K_{ij} + \partial_r (-\beta^r K_{ij} + \alpha \lambda_{ij}^r) = \alpha S_{ij},
\end{equation}
\begin{multline}
\lambda_{ij}^k = D^k_{ij} + \frac{1}{2}\delta_i^k(A_j + 2V_j - D_{jr}^{\;\;r})+Ê\\ \frac{1}{2}\delta_j^k(A_i + 2V_i - D_{ir}^{\;\;r}) - \frac{V^k\gamma_{ij}}{2},
\end{multline}
\begin{multline}
S_{ij} = -R_{ij}^{(4)} - 2 K_i^kK_{kj} + \text{tr}K K_{ij} + \\Ê\frac{2}{\alpha}(K_{ir}B_j^{\;r}+K_{jr}B_i^{\;r}-K_{ij}\text{tr}(B))+2D_{ik}^{\;\;r}D_{rj}^{\;\;k} 
+2D_{jk}^{\;\;r}D_{ri}^{\;\;k} + \\\Gamma^k_{kr}\Gamma^r_{ij} - \Gamma^k_{ri}\Gamma_{kj}^{r} - (2D^{\;kr}_k-A^r) ( D_{ijr}+D_{jir}) + \\ÊA_i \left(V_j - \frac{D_{jk}^{\;\;k}}{2} \right) +
A_j\left(V_i - \frac{D_{ik}^{\;\;k}}{2}\right)-V^kD_{kij} + \\Ê\frac{\gamma_{ij}}{4}(-D_k^{\;rs}\Gamma^k_{rs}+D_{kr}^{\;\;r}D_{s}^{ks}-2V^kA_k + \\Ê\text{tr}(K^2)-(\text{tr}K)^2+2 \alpha^2 G^{00}),
\end{multline}
\begin{equation}
R_{ij}^{(4)} = G_{ij} - \frac{\gamma_{ij}}{2}(-\alpha^2 G^{00}+\text{tr}(G)),
\end{equation}
\\\linebreak\textit{\underline{Momentum related variable:}}
\begin{multline}
\partial_t V_k + \partial_r (-\beta^r V_k + B^r_{\;i}-B_i^{\;r}) = \Omega_k = \\ \alpha[\alpha G^0_k+ A_r (K^r_k-\text{tr}(K)\delta^r_k)+(D_{kr}^{\;\;s}-2D_{rk}^{\;\;s})K^r_s \\ - K^r_k(D_{rs}^{\;\;s}-2D^{\;\;j}_{jr})] + 2(B_k^{\;r} - \delta_k^r \text{tr}B)V_r +2(D_{rk}^{\;\;s}-\delta^s_kD^j_{jr})B^r_{\;s},
\label{Vsor}
\end{multline}
\\\linebreak\textit{\underline{First order derivatives:}}
\begin{equation}
A_k = \partial_k \ln(\alpha), \;\;\;\; D_{kij} = \partial_k \gamma_{ij}/2.
\end{equation}
where $R_{ij}^{(4)}$ are the space components of the four-dimensional Ricci tensor. Although in the original formulation of Bona-Masso equations, $A_i$ and $D_{ijk}$ are considered independent variables with their own respective evolution equations, we have evolved the slicing and the metric first and then taken the derivatives. During the simulations we observed that, if they are evolved independently, then the derivative of the slicing (the metric) does not match with $A_i$ ($D_{ijk}$) which leads to errors. Therefore we will calculate them directly through $\alpha$ and $\gamma_{ij}$.

\subsection{Gauge choices}

The $3+1$ formulation does not define slice and lapse evolution uniquely, which gives us a freedom to choose the appropriate slicing and lapse depending on the system. The simplest slicing would be the geodesic slicing $Q = 0$, where the time component of the four metric stays constant throughout the simulation. However in some situations this slicing fails to support the stability of the program, as the evolution of the metric is coupled to it. For example, for black holes the geodesic slicing leads to a negative metric, thus to a numerically unstable algorithm \cite{brugmann1996adaptive}. Other slicings are: the maximal slicing $\text{tr}(K) = 0$, harmonic slicing $Q = \text{tr}(K)$, and ``$1+log$" slicing $Q = \text{tr}(K)/\alpha$. The maximal and $1+log$ slicings are also known as singularity avoiding slicings \cite{alcubierre2008introduction} because they collapse the time evolution close to the singularity. The harmonic slicing corresponds to a slicing where $\alpha$ fulfils the wave equation $(\partial_t^2 - \partial_x^2)\alpha = 0$ \cite{geyer1995slicing}. In the following tests we only use the geodesic and harmonic slicings and work with zero shift ($Q_i = 0$). 

The tests have periodic boundary conditions. We will avoid tests that include other boundary conditions, e.g. ``single Schwarzschild black hole", since they are known to be very challenging and are currently under intense research (combined with appropriate slicings and lapses).

\section{Lattice Boltzmann Model}

\begin{figure}[t]
 \center
        \includegraphics[scale=0.4, trim = 40mm 70mm 40mm 35mm, clip]{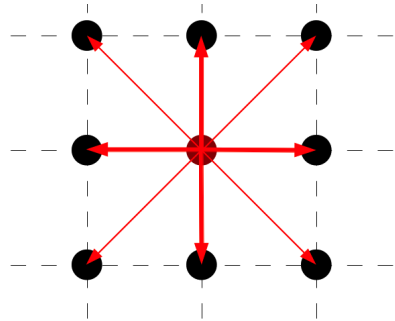}
        \caption{D2Q9 lattice configuration vectors. The thickness of the arrows represent the weights. The dark red point in the middle denotes the $(0,0,0)$ vector.}
        \label{d2q9}
\end{figure}

The lattice Boltzmann Method can be used to simulate fluids or solve partial differential equations in the form of conservation laws \cite{peng2011lattice}. We start with the discrete Boltzmann equation for the distribution functions, $f_\lambda$, 
\begin{equation}\label{lbe}
f_\lambda(x+c_\lambda \delta t,t+\delta t) - f_\lambda(x,t) = -\frac{f_\lambda(x,t)-f_\lambda^{eq}(x,t)}{\tau/\delta t},
\end{equation}
where $\tau$ is the relaxation time, and the index $\lambda$ represents the discrete velocity $\{ c_\lambda \}_{\lambda \in N}$. In this work we will use the Bhatnagar-Gross-Krook (BGK) approximation \cite{peng2011lattice}, which is a small amplitude approximation for the collision term of the lattice Boltzmann equation (rhs of Eq.~\eqref{lbe}). In case of an ideal gas, one can take $f^{eq}$ as the Maxwell-Boltzmann (MB) distribution, expanded in orthogonal polynomials, and recover the Navier-Stokes equations \cite{lbe}.
%The Lattice Boltzmann method is derived from the discretization of the Boltzmann equation and the main principles stem from Cellular Gas Automata \cite{cga}. We have a lattice consisting of $N$ discrete velocities  that travel on the lattice carrying the discrete distributions $f_\lambda$  \cite{discrete}. 
%The lattice is constructed from the minimization of entropy with respect to lattice spacing. It is shown that minimal entropy is achieved choosing the velocity magnitudes as the roots of Hermite polynomials \cite{lattice}. However the roots are not necessarily integers and it becomes difficult to create a lattice that does not coincide with the numerical grid (extrapolations would be necessary, which would lead to the accuracy loss that was avoided by choosing non integer spacings). When one obtains a lattice, one can calculate the appropriate weights for a quadrature which is necessary to recover the moments of the distribution function. For example, we know that the zeroth moment of the MB distribution gives the density and the first one the momentum density:
%\begin{equation}
%\rho = \int f_{MB} d^3v, \;\;\;\; \rho u_i = \int v_if_{MB} d^3v.
%\end{equation}
The macroscopic fields can be calculated by using the relations
\begin{equation}
\rho = \sum_\lambda f^\lambda, \;\;\;\; \rho u_i = \sum_\lambda c_{\lambda i} f^\lambda.
\end{equation}
which, in the case of fluid dynamics, $\rho$ is the fluid density and $\rho u_i$ the momentum density. 
%If we choose to expand our distribution function with respect to Hermite polynomials (which follows naturally from the entropic perspective) then the natural weights would be those of $w(v) = e^{-v^2}$ and we can make sure that the orthogonality conditions are fulfilled:
%\begin{multline}
%\int w(v)H_n(v)H_m(v) d^3v = N \delta_{nm} \to \\ \sum_\lambda w(c_\lambda) H_n(c_{\lambda}) H_m(c_{\lambda}) = N \delta_{nm}.
%\end{multline}
%$N$ is the normalization factor, which has different values for different choices of Hermite polynomials. The expansion of a distribution function w.r.t. Hermite polynomials is given explicitly by:
%\begin{multline}
%f_\lambda = w(c_\lambda)\sum_n \frac{1}{n!}a_n(\rho,u) H_n(c_\lambda) \\ \text{with} \;\; a_n = \frac{1}{N} \int f(\rho,v)H_n(v) d^3v.
%\end{multline}
%%\subsubsection{Lattice}
The lattices are described by their dimensionality (D) and amount of discrete velocity vectors (Q). An $n$-dimensional lattice with $m$ velocities would be denoted by D$n$Q$m$. Here, we will use D2Q9 and D3Q19 in two and three dimensions, respectively. The D2Q9 lattice configuration is given by:
\begin{equation}
(c_{0},c_{1,2},c_{3,4},c_{5,6},c_{7,8}) =
 \begin{pmatrix}
  0 & \pm 1 & 0 & \pm 1 & \pm 1\\
  0 & 0 & \pm 1 & \mp 1 & \pm 1\\
 \end{pmatrix}
 \end{equation}
where the weights are defined by $w_0 = \frac{4}{9}$, $w_{1,2,3,4} = \frac{1}{9}$, $w_{5,6,7,8} = \frac{1}{36}$, and the speed of sound $c_s = \frac{1}{\sqrt{3}}$ (see figure \ref{d2q9}).
On the other hand, the D3Q19 lattice configuration is given by:
 \begin{equation}
 \begin{pmatrix}
  0 & \pm 1 & 0 & 0 & \pm 1 & \pm 1 & \pm 1 & \pm 1 & 0 & 0\\
  0 & 0 & \pm 1 & 0 & \mp 1 & \pm 1 & 0 & 0 & \pm 1 & \pm 1\\
  0 & 0 & 0 & \pm 1 & 0 & 0 & \mp 1 & \pm 1 & \mp 1 & \pm 1\\
 \end{pmatrix}
 \end{equation}
with weights $w_0 = \frac{1}{3}$, $w_{1,2,3,4,5,6} = \frac{1}{18}$, $w_{\geq 7} = \frac{1}{36}$ and $c_s = \frac{1}{\sqrt{3}}$. Both lattices are accurate up to second order, which means that one can recover the moments of the distribution up to second order \cite{chikatamarla2009lattices}.

The discrete Boltzmann equation with a source term distribution $S_\lambda (x,t)$ is given by \cite{buick2000gravity}:
\begin{eqnarray}
\begin{aligned}
f_\lambda(x+c_\lambda \delta t,t+\delta t) - f_\lambda(x,t) = &-\frac{f_\lambda(x,t)-f_\lambda^{eq}(x,t)}{\tau/\delta t} \\ &+ \delta t\; S_\lambda (x,t).
\label{numlb}
\end{aligned}
\end{eqnarray}
%and
%\begin{equation}
%\partial_t f_\lambda(x,t) + c_\lambda \partial_x f_\lambda(x,t) = -\frac{f_\lambda(x,t)-f_\lambda^{eq}(x,t)}{\tau} + S_\lambda (x,t) \delta t.
%\label{momeq}
%\end{equation}
This equation describes the evolution of the distribution functions. 
%The equation \ref{momeq} describes the evolution of moments of the distribution to the first order. 
%By using the Chapman-Enskog expansion \cite{lolo,disclat},  
%We will see how the moments are evolving to the second order by using Chapman-Enskog expansion in the following section.
%\subsubsection{Equilibrium Distributions and Sources} 
We calculate the equilibrium and source distributions for each component of $\alpha$, $\gamma_{ij}$, $V_i$, and $K_{ij}$, such that the correct moments of the equilibrium distribution are satisfied, and consequently, the right macroscopic differential equations are recovered. Thus, one gets 
\begin{equation}
^\alpha f_{\lambda}^{eq} = w_{\lambda}\alpha,
\label{alfa}
\end{equation}
\begin{equation}
^\gamma f_{ij\lambda}^{eq} = w_{\lambda}\gamma_{ij},
\end{equation}
\begin{equation}
^Vf_{k\lambda}^{eq} = w_{\lambda}\left[V_{k}\left(1-\frac{c_{\lambda}^l\beta^l}{c_s^2}\right)+\frac{c_{\lambda}^i(B_{ki}-B_{ik})}{c_s^2} \right],
\end{equation}
\begin{equation}
^Kf_{ij\lambda}^{eq} = w_{\lambda}\left[K_{ij}\left(1-\frac{c_{\lambda}^l\beta^l}{c_s^2}\right)+\frac{\alpha c_{\lambda}^m\lambda_{ij}^m}{c_s^2}\right],
\label{ekst}
\end{equation}
\begin{equation}
^\alpha S_{\lambda} = w_{\lambda}\left(1-\frac{1}{2\tau}\right)\alpha (\beta^r A_r - \alpha Q),
\end{equation}
\begin{equation}
^\gamma S_{ij\lambda} =  w_{\lambda}\left(2-\frac{1}{\tau}\right)(\beta^r D_{rij} - \alpha (K_{ij}-s_{ij})),
\end{equation}
\begin{equation}
^VS_{i\lambda} =  w_{\lambda}\left(1-\frac{1}{2\tau}\right)\Omega_i,
\end{equation}
\begin{equation}
^KS_{ij\lambda} = w_{\lambda}\alpha S_{ij}\left(1-\frac{1}{2\tau}\right),
\label{ekstsor}
\end{equation}
where the symbol $^*$ at the left of the distribution and source term, $^* f$ and $^* S$, denotes the field to which $f$ and $S$ are associated. The macroscopic variables of concern are calculated by
\begin{equation}
\rho^* = \sum_\lambda {^*f}_\lambda + \frac{ \delta t ^*S}{2}.
\end{equation}
where $\rho^*$ stands for ($\alpha$, $\gamma_{ij}$, $V_i$, $K_{ij}$). By performing the Chapman-Enskog expansion \cite{buick2000gravity,guo2002discrete}, we see that the distributions and source terms recover the Bona-Masso formulation of Einstein equations to the first order in Knudsen number \cite{buick2000gravity,guo2002discrete}. One last remark is that we could fix the second order moment of the equilibrium distributions to zero, but it decreases the stability of the system for two reasons: first, the equilibrium distribution can become easier negative, which violates the H-theorem leading to an unstable evolution of the system; and second, the numerical viscosity increases the stability of the system in the same fashion as the numerical diffusivity, introduced in other methods, does it. We have finished with the model description. Now we will study some well-established examples in order to validate and characterise the method.

\section{Tests and Results}
\label{Tests and Results}

In order to validate our model, we will perform three numerical tests. The first two tests are flat space-times, in one we consider an ideal fluid and in the other one just vacuum for the energy-momentum tensor. Finally, the last test corresponds to a curved space-time in vacuum. The errors of the simulations are presented in section \ref{errors}.

\subsection{Expansion of flat universe}

\begin{figure}[h]
\center
        \includegraphics[scale=0.3, trim = 0mm 0mm 0mm 0mm, clip]{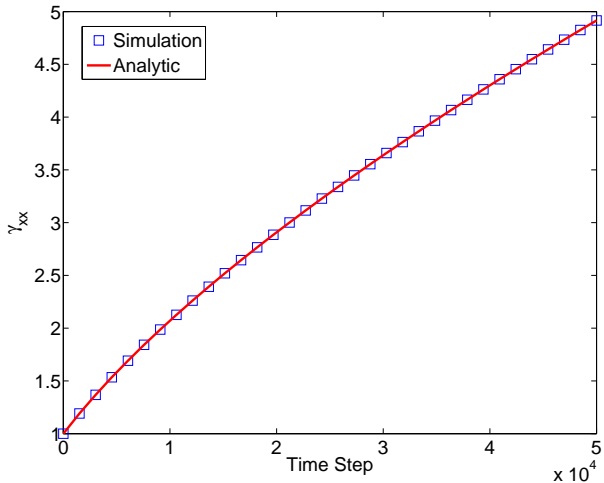}
        \caption{Time evolution of a single component of the metric tensor for an expanding universe. The metric starts from unity and evolves according to the time dependence given in Eq. \eqref{evolvmet}. The lattice size and spacing are irrelevant factors due to isotropy and flatness.}
        \label{gij}
\end{figure}

\begin{figure}[h]
\center
        \includegraphics[scale=0.3, trim = 0mm 0mm 0mm 0mm, clip]{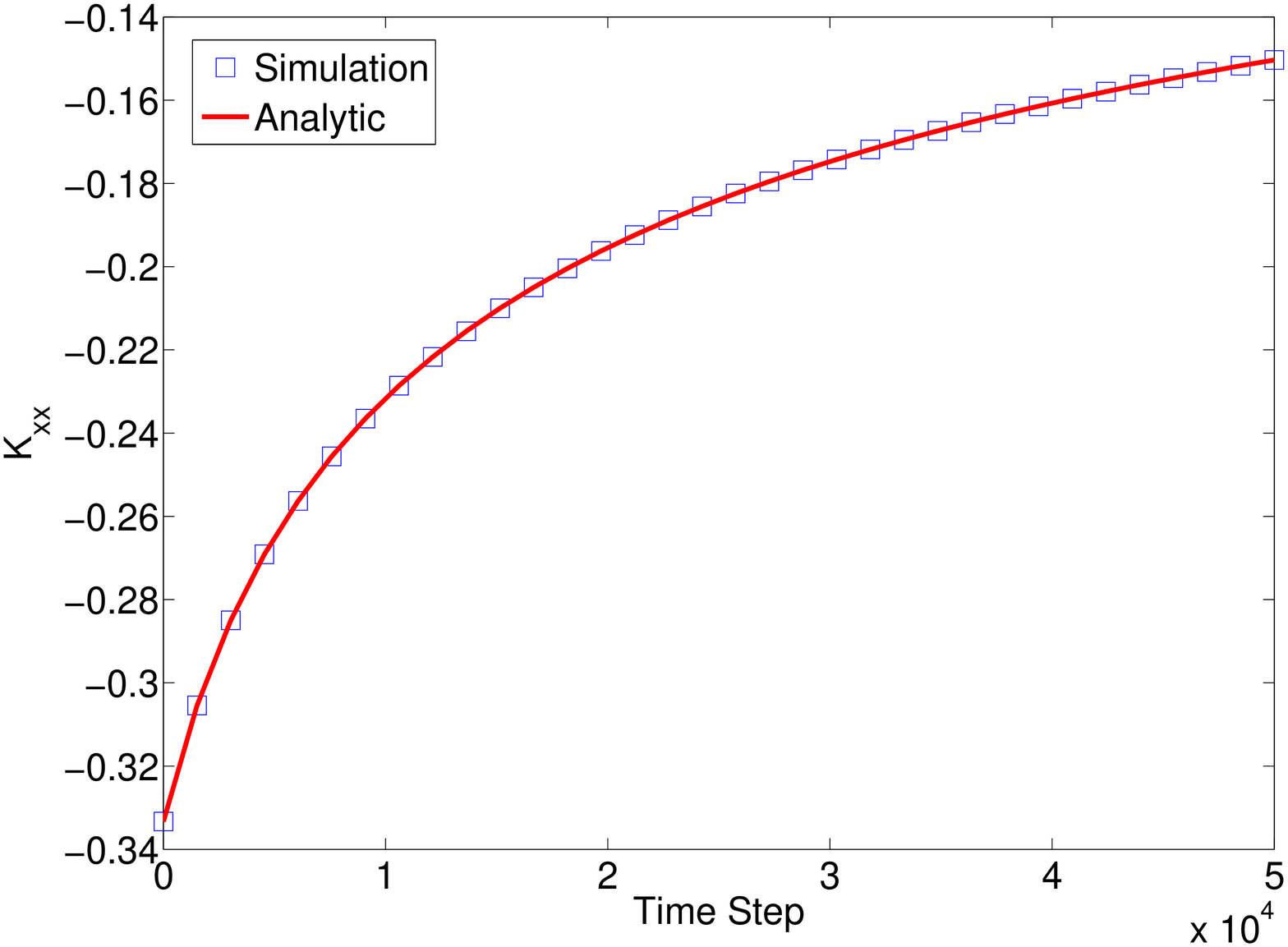}
        \caption{Single component of the second fundamental form, $K_{xx}$, for an expanding universe. This fundamental form describes the time derivative of the metric via $K(t) = -\dot{a}(t)/2$.}
        \label{kij}
\end{figure}

In this case, the space part of the energy-momentum tensor, $G_{ij}$, of an ideal fluid is given by
\begin{equation}
G_{00} = \rho(t)c^2, \;\;\;\;\;\; G_{ij} = P(t)\gamma_{ij},
\end{equation}
where $\rho(t)$ and $P(t)$ are the density and pressure of the fluid, respectively. By solving the Einstein equations for this system one obtains the Friedmann equations \cite{friedman1999curvature} (also knows as FLRW metric), which are
\begin{equation}
\frac{\dot{a}^2+kc^2}{a^2} = \frac{9 \pi G \rho + \Lambda c^2}{3},
\end{equation}
\begin{equation}
\frac{\ddot{a}}{a} = - \frac{4 \pi G}{3}\left(\rho + \frac{3P}{c^2}\right)+\frac{\Lambda c^2}{3},
\end{equation}
where $a$ is a quantity that determines the metric,
\begin{equation}
\gamma_{ij} = a(t)^2 \delta_{ij} = t^{\frac{2}{3}} \delta_{ij},
\label{evolvmet}
\end{equation}
and 
\begin{equation}
P(t) = \frac{1}{3t^2}.
\end{equation}
For an ideal gas with equation of state $P = \rho$ (assuming $c^2 = 8\pi G = 1$). The simulation ran using a D3Q19 lattice with $\delta t = \delta r = 0.001$, and a relaxation time $\tau/\delta t = 3$ (all values are given in numerical units). The results can be observed in Figs. \ref{gij} and \ref{kij}. We have used a lattice of $4\times 4\times 4$ cells due to the fact that the lattice size does not matter due to isotropy and homogeneity of the problem, i.e. the expansion is the same at every position. In Figs. \ref{gij} and \ref{kij}, we observe excellent agreement between the results of our simulation and the theory, showing that the universe expands following the Friedmann equations.
%\begin{figure}[h]
%\center
%        \includegraphics[scale=0.3, trim = 0mm 0mm 0mm 0mm, clip]{errorsRW.eps}
%        \caption{The relative error of the second fundamental form increases from a negative value to an asymptotic value around $2\cdot 10^{-5}$. }
%        \label{erk}
%\end{figure}
\subsection{Gauge Wave}

\begin{figure}[h]
\center
        \includegraphics[scale=0.3, trim = 0mm 0mm 0mm 0mm, clip]{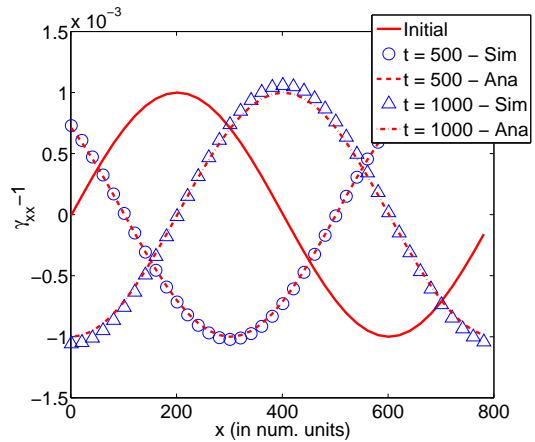}
        \caption{Evolution of the $xx$ component of the metric for the propagation of a gauge wave is shown at three different times. Here, $t$ denotes the numerical time step. We see that the simulation deviates slightly from the analytical values, but the shapes are preserved and the maximum (minimum) of the waves are on the same track. ``Ana" stands for analytical solution and ``Sim" for simulation.}
        \label{gijg}
\end{figure}

\begin{figure}[h]
\center
        \includegraphics[scale=0.3, trim = 0mm 0mm 0mm 0mm, clip]{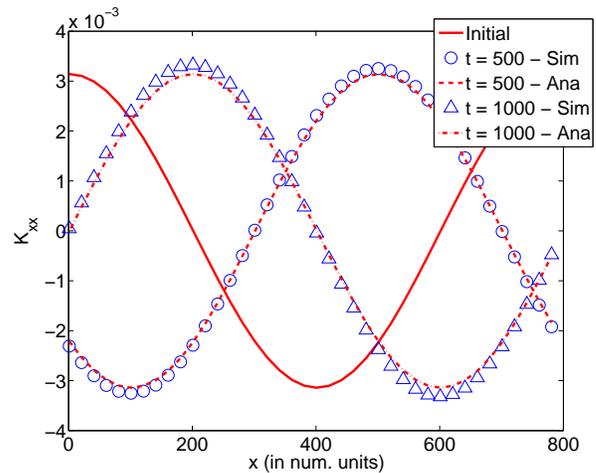}
        \caption{Evolution of the second fundamental form for the propagation of a gauge wave is shown at three different times. We see a similar behaviour to that of the metric. ``Ana" stands for analytical solution and ``Sim" for simulation.}
        \label{kijg}
\end{figure}

\begin{figure}[h]
\center
        \includegraphics[scale=0.3, trim = 0mm 0mm 0mm 0mm, clip]{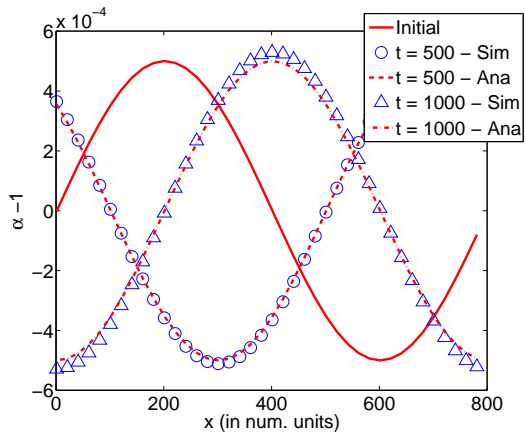}
        \caption{Evolution of $\alpha$ ($tt$ component of the metric) for the propagation of the gauge wave is shown at three different times. Here, we observe again a similar behaviour to that of the metric. ``Ana" stands for analytical solution and ``Sim" for simulation.}
        \label{erra}
\end{figure}

In this test, the metric tensor is given by
\begin{equation}
ds^2 = g_{\alpha \beta}dx^\alpha dx^\beta = -Hdt^2 + H dx^2 +dy^2 +dz^2,
\end{equation}
where $H = H(x-t) = 1 - A \sin\left(\frac{2\pi(x-t)}{d}\right)$ with $d$ the wavelength of the gauge wave, which in our case is set to unity, and $A = 10^{-3}$ its amplitude. The extrinsic curvature can be calculated by directly taking the time derivative of the three metric and dividing by $-2\alpha$,
\begin{equation}
K_{xx} = -\frac{\pi A}{d}\frac{\cos\left(\frac{2\pi(x-t)}{d}\right)}{\sqrt{H}}, \;\;\;\; K_{ij} = 0.
\end{equation}
It can be checked by Eq. \eqref{secfev} that the extrinsic curvature also evolves with the time derivative of this expression. The time evolution of $\alpha$ is given by
\begin{equation}
\partial_t \alpha = -\alpha^2Q \implies Q = \text{tr}(K).
\end{equation}
Here, we have chosen the harmonic slicing because we want $\alpha = \sqrt{H}$ to propagate also as a wave and the harmonic slicing provides a wave-like evolution to $\alpha$, while keeping the evolution of the quantity $H$ consistent with the evolution of the three metric. If the evolution of $\alpha$ and $\gamma$ do not lead to the same equations for $H$, then the respective time derivatives of the extrinsic curvature do not match. The results, using $\tau/\delta t = 4.5$, are presented in Figs. \ref{gijg}, \ref{kijg}, and \ref{erra}. We see again that they are in good agreement within a relative error of $1$\%. The simulations ran using a D2Q9 lattice configuration, where we have redefined the spatial coordinate $x \rightarrow x_r$ and keep $x$ for the positions in the lattice, being $x_r = -400 \delta r + \left(x - 1/2\right) \delta r$, and $\delta t = \delta r = 0.00125$.


\subsection{Linear Wave}

\begin{figure}[h]
	\center
        \includegraphics[scale=0.3, trim = 0mm 0mm 0mm 0mm, clip]{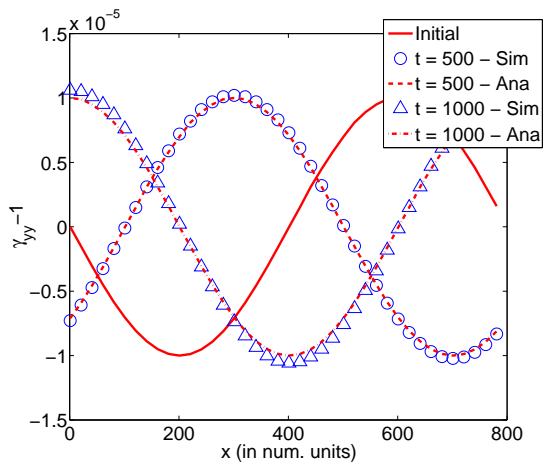}
        \caption{Evolution of the $xx$ component of the metric for the propagation of a linear wave is shown at three different times $t$. Note that the linear wave evolves similarly to the gauge wave.}
        \label{gyyshift}
\end{figure}

In this test, we set the following metric tensor:
\begin{equation}
ds^2 = g_{\alpha \beta}dx^\alpha dx^\beta = -dt^2 + dx^2 + (1+b)dy^2 + (1-b)dz^2,
\end{equation}
where $b = A \sin\left(\frac{2 \pi (x-t)}{d}\right)$ with $d$ the system size as above. The extrinsic curvature can be calculated directly by taking the time derivative of the three metric and dividing by $-2\alpha$, with $\alpha = 1$:
\begin{equation}
K_{yy} = \frac{\pi A}{d}\cos\left(\frac{2\pi(x-t)}{d}\right),
\end{equation}
\begin{equation}
K_{zz} = -\frac{\pi A}{d}\cos\left(\frac{2\pi(x-t)}{d}\right).
\end{equation}
In this case, we take geodesic slicing $Q = 0$, which would impose into the system the analytical evolution of $\alpha$. As the perturbation is traceless, geodesic slicing and harmonic slicing are identical gauges. The amplitude is chosen as $A = 10^{-5}$. For this simulation, we use a D3Q19 lattice configuration, $\tau/\delta t = 4.5$, and the same values for $x$, $t$, and $\delta t$, $\delta r$, as before. Note that the results are very similar to those of the gauge wave as expected (see Fig.~\ref{gyyshift}). 

%\begin{figure}[h]
%\center
%        \includegraphics[scale=0.3, trim = 0mm 0mm 0mm 0mm, clip]{relerrllog.eps}
%        \caption{The evolution of error of $\gamma_{yy}$ depending on $\tau$. A similar behavior to that of the gauge wave is observed. System crashes earlier with too small or too big $\tau$ and an optimal value for $\tau$, which gives enough stability for the longest time, is obtained.}
%        \label{gijl}
%\end{figure}	

\section{Analysis and Simulation Errors}
\label{errors}

In this section, we analyse the errors of the performed simulations of the previous section. The tests will be analysed in groups with the corresponding errors.

\subsection{Errors in the Validation Tests}

\begin{figure}[h]
\center
        \includegraphics[scale=0.26, trim = 0mm 0mm 0mm 5mm, clip]{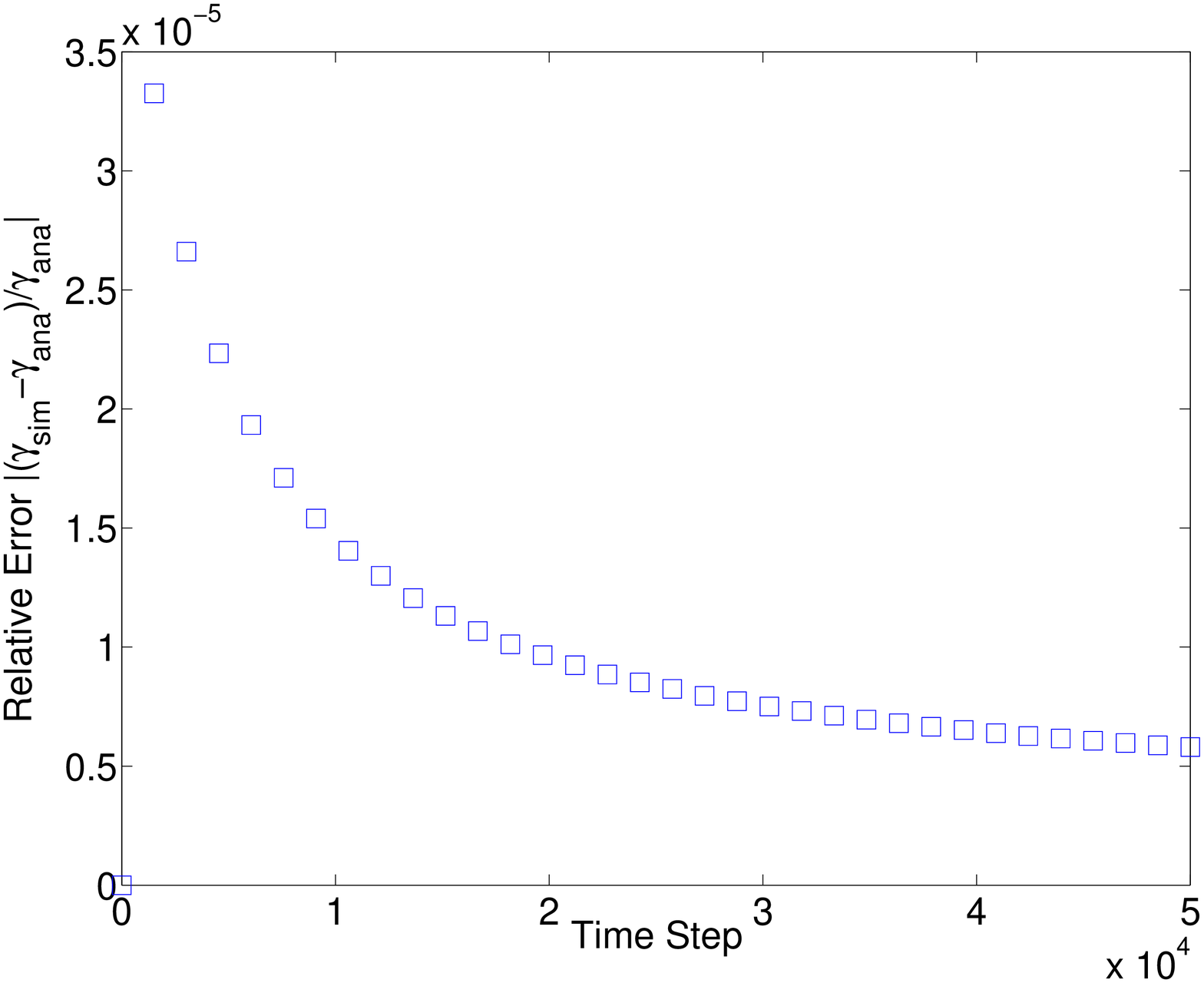}
        \vspace{-5mm}
        \caption{Relative error of the metric for an expanding universe. It is observed that throughout the simulation, it is always below $0.01$\%.}
        \label{erg}
\end{figure}

The relative error of the metric for the case of an expanding flat universe is shown in Fig.~\ref{erg}. As it can be seen, the error decreases monotonically while the simulation stays in good agreement with the theory.

On the other hand, the gauge and linear waves show very similar behaviour in our simulations. The test of the gauge wave is expected to get unstable by design due to emerging singularities. As mentioned in Ref.~\cite{alcubierre2004towards}, such system with $T^3$ topology must have a singularity in the future or in the past and its effect must show at some time during the simulation. Our simulation ran for almost $2 \cdot 10^{3}$ time steps till the relative error exceeds $10$\%. Compared to the well-developed methods of ADM and BSSNOK with the use of Cactus or other numerical solvers, which can handle much more time steps, our model needs improvements. This error introduced by the lattice Boltzmann model is due to the fact that singularities possess large gradients in the geometric variables and they lead to negative equilibrium distribution functions, and consequently, to numerical instabilities. Thus, we can expect that by adding the H-theorem, i.e. introducing a entropic lattice Boltzmann model \cite{karlin1999perfect, boghosian2003galilean}, one can improve drastically the model keeping its simplicity.
%This test is expected to have an accumulated phase error in addition to the numerical error which we presented in figure \ref{gyyshift}. For general comparison purposes, we monitored the Hamiltonian constraint during the simulation of a linear wave (Fig. \ref{haml}). As the gauge wave had only $K_{xx}$ and $\Gamma_{xx}^x$ inequal to zero, the Hamiltonian constraint was fulfilled trivially. In the case of a linear wave, we have a curved space wave, which gives a more complicated expression for the Hamiltonian constraint. We see that the violation of the Hamiltonian constraint start to increase rapidly around the time step $5 \cdot 10^{3}$. The Hamiltonian constraint is shown to be a good stability parameter of the simulation.

\subsection{Relaxation Time $\tau$ and Numerical Diffusivity $\eta$}

The relaxation time $\tau$ is one of the characteristic parameters of any single relaxation lattice Boltzmann scheme. More precisely, it is related to the numerical diffusivity through 
\begin{equation}
\eta =  \left(\tau-\frac{\delta t}{2} \right) c_s^2 .
\label{numdifrel}
\end{equation}

\begin{figure}[h]
\center
        \includegraphics[scale=0.3, trim = 0mm 0mm 0mm 0mm, clip]{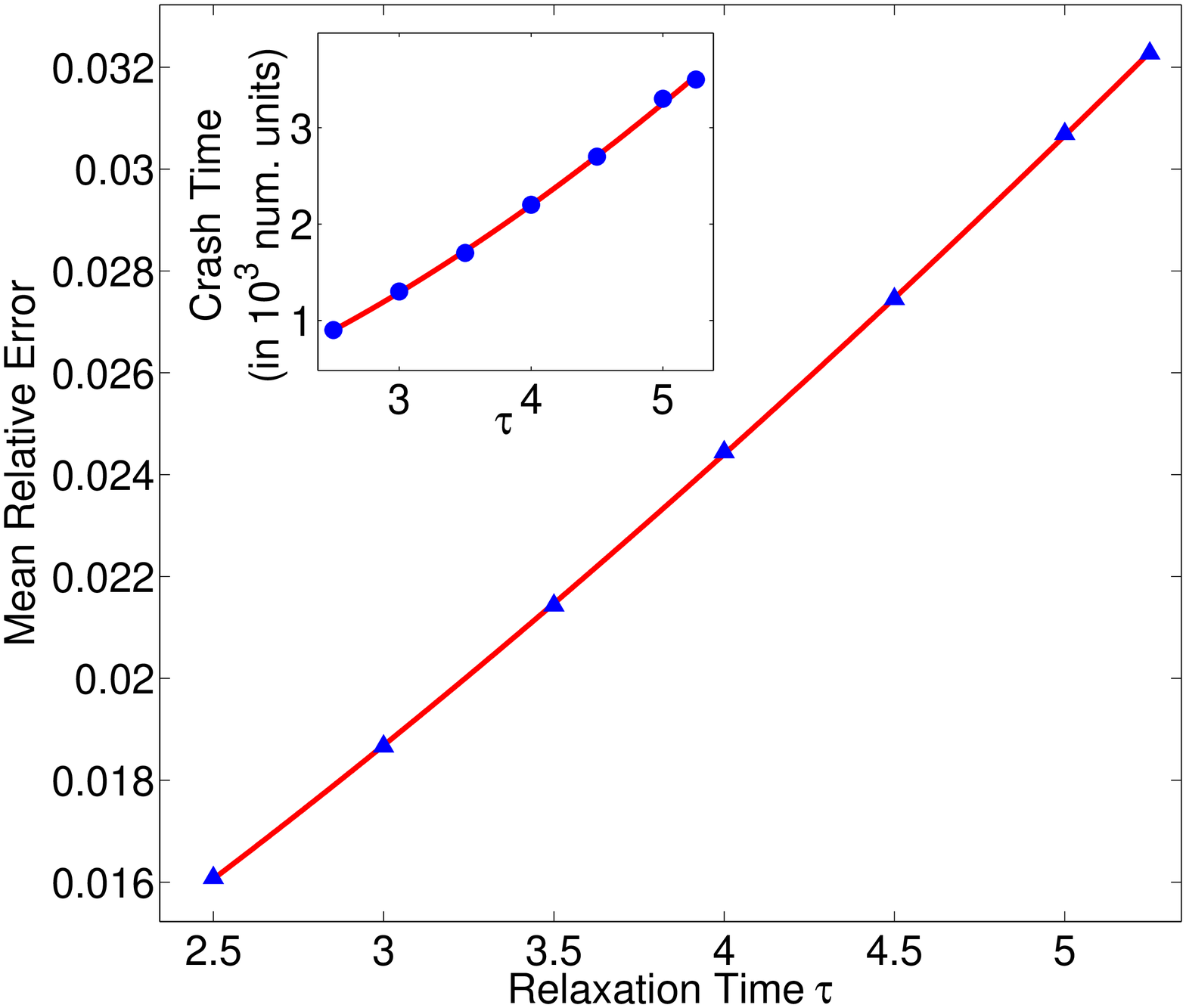}
        \vspace{-5mm}
        \caption{Mean relative error of $\gamma_{xx}$ for the propagation of gravitational waves at the first $500$ time steps. For higher relaxation time the accuracy gets worse. However, it can be seen, in the inset, that the life-time of the simulation increases with increasing the relaxation time.}
        \vspace{-5mm}
        \label{gijl}
\end{figure}

For the above tests, $\tau/\delta t$ varies between $2.5$ and $5.25$, and consequently, $\eta$ takes values between $0.0008\bar{3}$ and $0.00208\bar{3}$. We see that as the relaxation time increases, the stability of the system also increases (see inset of Fig.~\ref{gijl}), while the errors become larger (see Fig.~\ref{gijl}). For instance, we observe that while $\tau/\delta t = 2.5$ leads to numerical instabilities quicker than $\tau/\delta t = 4.5$, it also introduces a less mean relative error into the system. Indeed, further tests show that there is an optimal value of $\tau$ for which the error introduced by the model is less while the stability is better. In the case of gravitational waves $\tau/\delta t = 4.5$ is the optimal relaxation time.

\begin{figure}[h]
\center
\includegraphics[scale=0.3,trim = 0mm 0mm 0mm 0mm, clip]{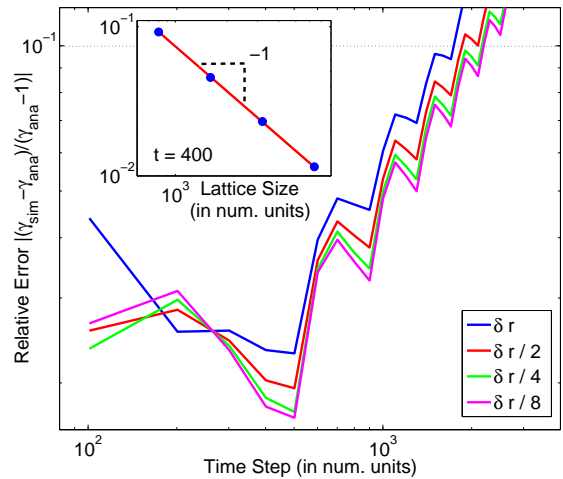}
\vspace{-5mm}
\caption{Main frame: relative error of the metric for the propagation of gravitational waves at different time steps for different resolutions. Here, we have kept constant the numerical diffusivity, $\eta = 0.001\bar{6}$. The errors strongly depend on the lattice spacing, and finer lattices lead to longer simulation time. Inset: we see that with finer lattices the relative error introduced at time step $400$ $\delta t$ decreases as a power law showing convergency.}
\vspace{-5mm}
\label{simana}
\end{figure}

Note that Eq.~\eqref{numdifrel} states that by increasing the relaxation time, the numerical diffusivity also increases for a fixed $\delta t = \delta x$, but one can also fix the numerical diffusivity while changing independently the relaxation time. Thus, in a second test (see Fig.~\ref{simana}) we study the behaviour of the error when one increases the relaxation time and decreases the lattice spacing, such that the numerical diffusivity remains constant. We observe that the simulation lasts about $50$\% longer in order to obtain the same accuracy. Therefore, we conclude that the numerical diffusivity tunes the accuracy, while the relaxation time the stability of the model. 

As a final error test, the inset of Fig.~\ref{simana} shows that the error decreases linearly with the lattice spacing (at time $t = 400 \delta t$), concluding that our model converges to the analytical solution when the lattice spacing is decreased.

%\begin{figure}[h]
%	\center
%        \includegraphics[scale=0.3, trim = 5mm 0mm 0mm 0mm, clip]{hamiltonian11.eps}
%        \caption{Evolution of the Hamiltonian constraint for $\tau = 11$. Hamiltonian is a parameter of the system which must be equal to zero all the time, implying energy conservation. Due to numerical errors, it is not possible to maintain an exact 0 value. However it is a good indicator of system stability \cite{lele,book}. We see that it is in accordance with the crash time of the metric.}
%        \label{haml}
%\end{figure}
%\subsection{Relative Error of $K_{ij}$}
%For the linear wave we present the relative error in $K_{ij}$. As $K_{ij}$ has a very small value, the relative error of it grows very quickly. However its impact on the metric is still relatively small. One of the reasons why $K_{ij}$ deviates from the analytical solution so quickly could be that it is the only quantity that has a non-zero flux, a non-zero first moment. This first moment could pose a problem in two ways: it grows too big and the simulation is out of its low order expansion region and the equilibrium distribution may get negative leading to an ill-defined entropy and an unstable system. 
%\begin{figure}[h]
%\center
%        \includegraphics[scale=0.3, trim = 0mm 0mm 0mm 0mm, clip]{relerrglogshort.eps}
%        \caption{The evolution of the error of $xx$ component of the metric depending on $\tau$. We see that for smaller $\tau$ values, the error introduced in the beginning is less. However for smaller times, the crash time is earlier. The points, where the lines end, are the crash points. We observe that after $\tau = 16$, higher $\tau$ values lead to higher errors which cross $10^{-2}$ limit earlier.}
%        \label{errg}
%\end{figure}
\section{Parallelisation of the code}

\begin{figure}[h]
\center
        \includegraphics[scale=0.3, trim = 5mm 5mm 5mm 5mm, clip]{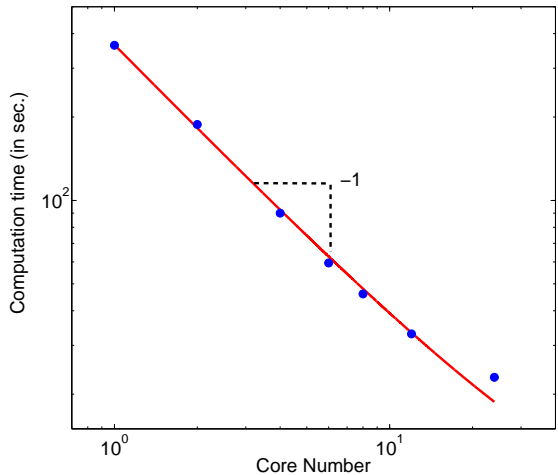}
        \vspace{-0mm}
        \caption{Computational time as function of the number of CPU cores. Here we observe that the computational time falls rapidly with increasing the number of CPU cores and is fitted very well by an inverse linear function.}
        \vspace{-5mm}
        \label{comptime}
\end{figure}

Our code has been parallelised with OpenMP and tested with $1-24$ cores. The results are given in the Fig.~\ref{comptime}. The linear decrease in total computational time with respect to number of cores shows that our model is optimal for parallelisation. Note that for $24$ cores, the total computational time deviates slightly from the linear behaviour because we have performed our simulations in nodes with a maximum of $24$ cores, which implies that we have reached the physical limit, and usually at this limit tasks are not performed very efficiently. Other more sophisticated implementations as MPI and CUDA, for GPUs, will be a subject of future works.

\section{Conclusions}
\label{Conclusions}
In this paper, we have developed a lattice Boltzmann model for solving Einstein equations, using the Bona-Masso formalism. We have validated our model with well-established tests, namely the expansion of a flat universe, and the propagation of gauge and linear waves. The expansion of a flat universe was recovered accurately and the wave tests showed good agreement with the analytical solutions. The roles of the relaxation time and the numerical diffusivity were also studied finding that they are crucial in determining the stability and accuracy of the model. In particular, we have observed that the system gains stability (accuracy) by increasing (decreasing) the relaxation time, and therefore, an optimal value, that compromises both, can be found. More precisely, the numerical diffusivity tunes the accuracy while the relaxation time the stability of the model. 

In addition to the validation tests, the inverse linear dependence of computational time with respect to the number of CPU cores was demonstrated, which is a major strength of lattice Boltzmann methods. It must be clearly underlined that with further work on this model, e.g. entropic extensions, the lattice Boltzmann method might offer new numerical advantages to Numerical Relativity.

\begin{acknowledgments}
We would like to thank Erik Schnetter and Ruxandra Bondarescu for bringing insights into numerical relativity and for their valuable suggestions. Additionally, we also thank Hans Herrmann for useful discussions. MM acknowledges financial support from the European Research Council (ERC) Advanced Grant 319968-FlowCCS.
\end{acknowledgments}

\bibliography{biblio}

\end{document}